\documentclass{raa}

\usepackage{graphicx,times}
\usepackage{natbib}
\usepackage{amssymb,amsmath}
\bibpunct{(}{)}{;}{a}{}{,}

\usepackage{multirow}
\usepackage{ulem}

\usepackage[pagebackref=true]{hyperref}
\hypersetup{%
    colorlinks=true,
    linkcolor=green,
    anchorcolor=red,
    citecolor=blue,
    filecolor=red,
    urlcolor=red}

\begin{document}
\title{Revealing Double White Dwarf Mergers with Multi-messenger Signals}

\volnopage{Vol.0 (20xx) No.0, 000--000}      
\setcounter{page}{1}          

\author{He-Wen Yang 
      \inst{1,2}
    \and Pak-Hin Thomas Tam 
      \inst{1,2}
    \and Lili Yang 
      \inst{1,2}
}
\institute{School of Physics and Astronomy, Sun Yat-sen University, Zhuhai 519082, China; {\it tanbxuan@mail.sysu.edu.cn,yanglli5@mail.sysu.edu.cn}\\
    \and
    CSST Science Center for the Guangdong-Hong Kong-Macau Greater Bay Area, Sun Yat-Sen University, Zhuhai 519082, China\\
\vs\no
    {\small Received 20xx month day; accepted 20xx month day}
}

\abstract{A significant number of double white dwarfs (DWDs) are believed to merge within the Hubble time due to the gravitational wave (GW) emission during their inspiraling phase. The outcome of a DWD system is either a type Ia Supernova as the double-degenerate model, or a massive, long-lasting merger remnant. Expected multi-messenger signals of these events will help us to distinguish detailed merging physical processes. In this work, we aim to provide a generic scenario of DWD merging, investigate the emission of all major messengers, with a focus on GWs and neutrinos. Our goal is to provide some guidance for current and future (collaborative) efforts of multi-messenger observations. Throughout the merging evolution of a DWD system, different messengers (GW, neutrino and electromagnetic wave) will dominate at different times. In this work, we show that DWD merger events located at the distance of $1\ \mathrm{kpc}$ can indeed produce detectable signals of GWs and neutrinos. The GW frequency are in $0.3-0.6\ \mathrm{Hz}$ band around $10$ days before tidal disruption begin. We estimate that in optimistic situations, the neutrino number detected by upcoming detectors such as JUNO and Hyper-Kamiokande can reach O(1) for a DWD merging event at $\sim$1~kpc.
}

\authorrunning{H.-W. Yang, P.H.T. Tam \& L. Yang}            
\titlerunning{GWs and neutrinos from DWDs}  

\maketitle

\section{Introduction}
Many evidences show that most stars in the Universe are not single isolated stars, but in binary systems~(\citealt{Sana2012_Binary,Duch^ene2013,El-Badry2021}). Stellar evolution models suggest that most stars will become white dwarfs (WDs). Therefore, the most common compact object mergers form out of double white dwarf (DWD) binaries.

In the past few years, some studies of DWD mergers show that every close, DWD binary may eventually merge due to energy loss and angular momentum loss through gravitational wave (GW) radiation~(\citealt{Shen2015}). DWD mergers can lead to various kinds of observable astronomical phenomena including some most luminous transient events in the Universe. On one hand, DWDs are treated as one of the progenitors of type Ia Supernovae (SNe Ia) for a long time, known as the double-degenerate (DD) scenario~(\citealt{Yungelson2016,Livio2018}). Other suggested manifestations of DWD mergers are transient electromagnetic signals such as fast blue optical transients~(\citealt{Lyutikov19}) and less-luminous nova-type explosions~(\citealt{Roy22}). On the other hand, if the merger core fails to ignite, the merger product is not clear, suggestions include a rapidly rotating WD with a strong magnetic field~(\citealt{Ji2013,Rueda2019}) which latter collapses to a neutron star (NS)~(\citealt{Ruiter2019,Liu2020}), or a heavy, long-lasting WD~(\citealt{Schwab21,WuCY22}). Therefore, a full understanding of the outcomes of DWD mergers not only provides us with the merger rates of the most common compact objects in the Universe, it will also help us to understand the origin of a broad range of transient phenomena and the formation of a class of specific WDs and NSs.

DWD systems are important gravitational wave sources~(\citealt{Nelemans01,Huang2020}). During the inspiral phase, DWD systems are stable, low frequency ($1-10\ \mathrm{mHz}$) GW sources, and are expected to be the first detected GW sources for the space-based GW observatories like LISA~(\citealt{Robson2019}), Taiji and TianQin~(\citealt{Gong21}). In the Galaxy, the high number distribution of low-mass (CO-He) DWDs makes them the dominant GW sources at frequency $\ge5\ \mathrm{mHz}$, while massive (CO-CO or ONeMg) DWDs are less numerous, and the former evolve to merge more quickly than the latter after common-envelope phases or stable Roche lobe overflow phase~(\citealt{Yu2010}). However, if they successfully evolve to very close binary (i.e., they tend to merge soon), the GW frequency of the system can reach $0.1\ \mathrm{Hz}$ or even $1\ \mathrm{Hz}$~(\citealt{Maselli2020,Zou2020}). 

Neutrinos are believed to be generated from astrophysical sources such as stellar cores, novae, core-collapse supernovae and compact binary mergers at $\mathrm{MeV}$ energies or from the cosmic rays accelerators at high and ultra-high energies~(\citealt{Halzen2002,Katz2012,Vitagliano2020}). However, the merging physics of DWDs is still unclear, and few works involving numerical simulations focus on neutrino emission. During the merging stage, the interaction region of DWD can create density and temperature conditions which are high enough for efficient neutrino 
production such that neutrinos can dominate the carried-away energy~(\citealt{Aznar-Sigu'an2014}). In recent years, some single-degenerate models indicate that “normal” SNe Ia bursts can produce significant neutrino flux that is four orders of magnitude smaller than that of core-collapse SNe, regardless of the exact explosion mechanism~(\citealt{Kunugise2007,Odrzywolek2011,Seitenzahl2015,Wright2017}). However, the neutrino flux associated with double-degenerate model has rarely been studied by simulations. It has previously been suggested that neutrino signals could help to distinguish these two progenitor models of SNe Ia~(\citealt{Raj2020}). It is commonly thought that SNe Ia come from the detonation of WD carbon cores when the temperature and density reach the ignition condition. Besides, the neutrino production is efficient when the temperature reaches above $10^{9}\ \mathrm{K}$, and carbon burning can reach such temperature~(\citealt{Itoh1996}), and in such case neutrino production will dominate the energy release.

In this work, we investigate the multi-messenger detection prospects of DWDs, focusing on gravitational waves and neutrinos. We mainly focus on the CO DWD systems which are detached binaries. In Section \ref{sec:mec}, we provide a simple description of DWD merging processes and a schematic overview of possible multi-messenger signals as well as their emission times for a merging event. In Section \ref{sec:mul}, we assume a toy model for gravitational wave and neutrino production when DWDs merge, introducing some physical quantities to describe the two types of signals. In Section \ref{sec:det}, we calculate the amplitude spectral density for the final inspiraling evolution and the neutrino number events from the merging process, and discuss their detectability using current and planned detectors. Finally, we present the discussions and conclusion of our results in Section \ref{sec:dis&con}.

\section{The mechanism of double white dwarf mergers}
\label{sec:mec}

\subsection{Physical processes}

In recent years, the merging dynamics of DWDs have been investigated by smoothed particle hydrodynamics (SPH) simulations, which reveal the details of the merging processes~(\citealt{Lor'en-Aguilar2009,Pakmor2010,Pakmor_2012ApJ,Raskin2012,Ji2013,Sato_2015,Sato_2016ApJ}). In general, it is thought that the less massive WD could be tidally disrupted by the more massive WD that remains. A small amount of mass of the disrupted one, which carries most of the angular momentum, are fastly accreted onto the surface of the undisrupted one, forming a thick Keplerian disk.

Depending on the initial masses and mass transfer rate of the DWD, it may result in the immediate ignition of the carbon core of the more massive WD that explodes (a violent merger), or that the product fails to be detonated after they merge. In the latter case, a magnetized corona above the disk and a strongly magnetized bi-conical jets perpendicular to the disk can form because of the development of the magneto-rotational instability within the disk~(\citealt{Ji2013}). And a viscous outflow is driven at the interface of the corona and the jet. It is also argued that the merging may go through some viscous processes with a timescale longer than the dynamical timescale, producing SN Ia explosion. Otherwise, if there is no explosion, the merge product is a more massive (ONeMg) WD. Even when the remnant mass exceeds the Chandrasekhar limit, the remnant will undergo a long-term evolution before it finally collapses to form a NS~(\citealt{Schwab_2021}).

\subsection{Multi-messenger Signals}

\begin{figure}[ht!]
    \centering
    \includegraphics[width=\hsize]{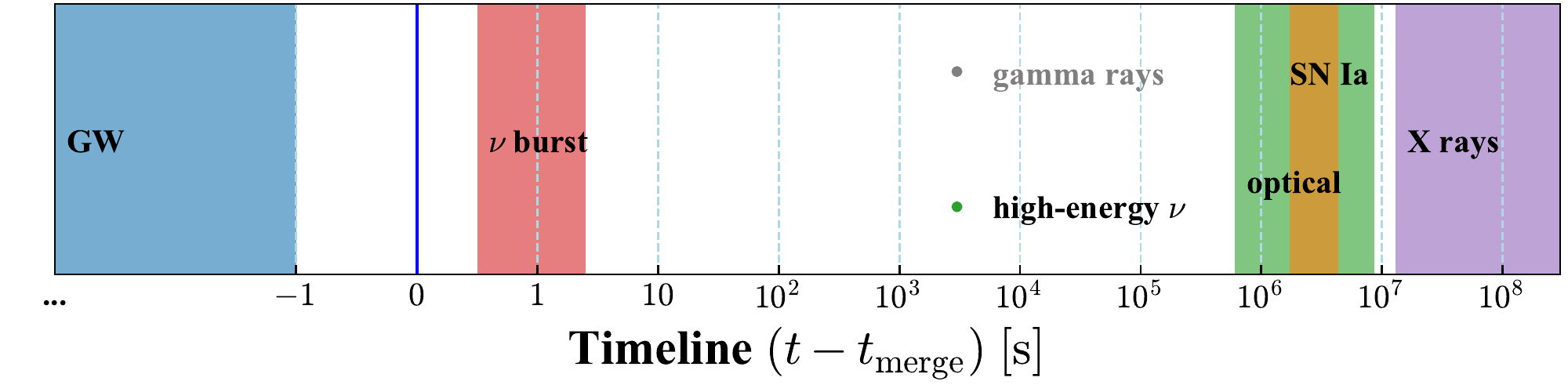}
    \caption{Plausible signals timeline before and after DWD merging in a multi-messenger perspective}
    \label{fig:timeline}
\end{figure}

In Figure \ref{fig:timeline}, we depict anticipated emission signals from DWD merger events. A plausible timeline of the signals associated with various processes is outlined as follows:

\begin{itemize}
    \item During the inspiral phase, the emission is dominated by GW. It is believed that GW emission reaches a maximum when tidal disruption occurs and thereafter is largely suppressed. 
    \item When tidal disruption starts, the dynamical timescale is thought to be $\sim1\ \mathrm{s}$ as estimated by the free-fall motion of the accreting mass across the tidal disruption radius of the secondary.
    \item Subsequently, a neutrino burst with a time duration of $\mathcal{O}(1\ \mathrm{s})$ follows immediately the dynamical merging process. 
    \item If SNe Ia explosion occurs, most materials of the DWD is thrown out, giving rise to a supernova light curves that may differ from (and indeed are likely broader than) those in  Chandrasekhar-mass explosion models~(\citealt{Fryer2010,Moll2014}). The luminosity reaches the maximum at $\sim 30$ days after they merge. 
    \item If there is no explosion, then in the viscous outflow process, cosmic rays are being accelerated via magnetic dissipation in the outflow region, subsequently leading to high-energy neutrino emission $\sim10^4\ \mathrm{s}$ after the merge~(\citealt{Xiao_2016ApJ}). It is also argued that gamma rays might be produced too, but can hardly be detected because of the surrounding optically thick environment. About $\sim 7$ days, after the merge optical photons are emitted due to the adiabatic cooling of ejecta. In addition, fallback accretion and spindown of the newly-formed central WD will produce X-rays about $150-200$ days post-merging~(\citealt{Rueda2019}).
\end{itemize}

\subsection{Merger Rate}

The merger rate of DWDs is the highest among all types of merger events in the Universe. In the Milky Way, the merger rate of DWD is expected to be $(1-80)\times10^{-13}\ \mathrm{yr}^{-1}M_\odot^{-1} \ (2\sigma)$ or $(7\pm2)\times10^{-13}\ \mathrm{yr}^{-1}M_\odot^{-1} \ (1\sigma)$~(\citealt{10.1093/mnras/stx102}). For comparison, by using the Milky Way stellar mass $6.4\times10^{10}\ M_\odot$ and the extrapolating factor of Milky Way-type galaxies $1.16\times10^{-2}\ \mathrm{Mpc}^{-3}$~(\citealt{Kalogera2001}) for DWDs, the merger rate of double neutron stars (DNSs) and double black holes (DBHs) are shown in Table~\ref{tab:megerrate}, which are smaller than DWDs by $2-4$ orders of magnitude. Additionally, previous studies also show that DWD merger rate are consistent and on the same order of SNe Ia rates deduced from observations~(\citealt{Toonen2012,Maoz2014,Maoz2018}).

\begin{table}
\bc
\begin{minipage}[]{100mm}
\caption[]{DWD, DNS and DBH merger rate in the local Universe and in the Milky Way
\label{tab:megerrate}}
\end{minipage}
\setlength{\tabcolsep}{1pt}
\small
    \begin{tabular}{cccc}
        \hline\noalign{\smallskip}
        {System} & {$\mathcal{R}\ [\mathrm{Gpc}^{-3}\ \mathrm{yr}^{-1}]$} & {MW Event Rate $[(100\ \mathrm{yr})^{-1}]$} & {References}  \\
        \hline\hline\noalign{\smallskip}
        DWD & $(0.74-5.94)\times10^{6}\ (2\sigma)$ & $0.64-51$ & \multirow{2}{*}{(\citealt{10.1093/mnras/stx102})}\\
        DWD & $(5.2\pm1.5)\times10^{5}\ (1\sigma)$ & $4.5\pm1.3$ & \\
        \hline\noalign{\smallskip}
        DNS &  $320^{+490}_{-240}$ & $2.8^{+4.2}_{-2.1}\times10^{-3}$ & (\citealt{Abbott_2021}) \\
        \hline\noalign{\smallskip}
        DBH &  $23.9^{+14.3}_{-8.6}$ & $2.1^{+1.2}_{-0.7}\times10^{-4}$ & (\citealt{Abbott_2021}) \\
        \noalign{\smallskip}\hline
    \end{tabular}
\ec
\end{table}

\section{A multi-messenger perspective of double white dwarf mergers}
\label{sec:mul}

In this section, we focus on the anticipated GW and neutrino emission from a DWD merging event.

\subsection{Gravitational Waves}
\label{subsec:GW}

In this work, we assume that there is no mass-transfer nor interaction between the two WDs until tidal disruption happens, because we assume that both components of the binary are of the type CO-WD or even more massive (ONeMg) WD. It is therefore reasonable to assume a two point-mass approximation during the inspiral phase.

Considering a DWD system with component masses $m_1$ and $m_2$ and an orbital separation of $a$, the orbital frequency is derived from its Keplerian motion as: 
\begin{align}\label{eqn:f_orb}
    f_\mathrm{orb} = \sqrt{\frac{G(m_1+m_2)}{2\pi^2a^3}},
\end{align}
where $G$ is the gravitational constant. In the case of a circular orbit, the GW is radiated only in the second harmonics, and hence the GW frequency $f$ is twice that of the orbital frequency $f = 2f_\mathrm{orb}$. During the inspiraling phase, the two independent polarization states (or waveforms) $+$ and $\times$ of GW radiation from DWD in the quadrupole approximation~(\citealt{Landau_1962}) is
\begin{align}
   h_+ &= \frac{\left(G\mathcal{M}_c\right)^{5/3}}{c^4 d}\left(\pi f\right)^{2/3}2(1+\cos^2{i})\cos(2\pi ft), \\
   h_\times &=\frac{\left(G\mathcal{M}_c\right)^{5/3}}{c^4 d}\left(\pi f\right)^{2/3}4\cos{i}\sin(2\pi ft).
\end{align}
where $c$ is the speed of light,$\mathcal{M}_c = (m_1m_2)^{3/5}/(m_1+m_2)^{1/5}$ is the chirp mass and $d$ is the distance to the source. The GW emission luminosity is:
\begin{align}
    L_\mathrm{GW}=\dot{E}=\frac{32}{5}\frac{G^4}{c^5}\frac{m_1^2m_2^2(m_1+m_2)}{a^5}.
\end{align}
Alternatively, it can be expressed as a function of chirp mass and GW frequency
\begin{align}
    L_\mathrm{GW}=\frac{32}{5}\frac{G^{7/3}}{c^5}\left(\pi f\mathcal{M}_c\right)^{10/3}.
\end{align}
Due to GW emission loss, the GW frequency evolves as:
\begin{align}
    \dot{f} = \frac{96}{5\pi}\frac{(G\mathcal{M}_c)^{5/3}}{c^5}(\pi f)^{11/3}.
\end{align}
Introducing the strain amplitude $h_0$ (or the instantaneous root-mean-square amplitude),
\begin{align}
    h_0 &= \sqrt{\left<h_+^2\right>+\left<h_\times^2\right>} =\sqrt{\frac{32}{5}}\frac{\left(G\mathcal{M}_c\right)^{5/3}}{c^4 d}\left(\pi f\right)^{2/3}.
\end{align}
Averaging over a full orbital period and all inclination angle, it is related to the GW luminosity by~(\citealt{Postnov_2014})
\begin{align}
    h_0^2 = \frac{1}{(\pi d)^2}\frac{G}{c^3}\frac{L_\mathrm{GW}}{f^2}.
\end{align}
The characteristic strain amplitude $h_c$ for inspiraling binaries is given by~(\citealt{Finn2000,Moore2014})
\begin{align}
    h_c^2 =\left(\frac{2f^2}{\dot{f}}\right)h_0^2 = \frac{2}{3}\frac{(G\mathcal{M}_c)^{5/3}}{c^3 d^2}f\left(\pi f\right)^{-4/3}.
\end{align}
The amplitude spectral density (ASD) is defined as $\sqrt{S_h}=h_cf^{-1/2}$, where $S_h$ is the power spectral density of sources~(\citealt{Moore2014}). Similarly, for GW detectors, $\sqrt{S_N(f)}$ is the effective strain spectral density of the detector noise used in sensitivity curves~(\citealt{Robson2019,Huang2020}).

Inspiraling of compact binaries can be well-modeled in theoretical analysis while the merging of binaries (especially those bursting sources) is poorly modeled~(\citealt{Moore2014}). As mentioned above, we consider the binary evolution during the inspiral phase, during which GW frequency is slowly changing $\Delta f = \dot{f} \Delta t$, and the GW signal is weak and remains almost unchanged for many years, then the corresponding signal is accumulated at detectors over a long time $\Delta t$, increasing the signal-to-noise ratio (SNR). The ASD can be described as $\sqrt{S_h} = \sqrt{2}h_{0}\sqrt{\Delta t}$. However, for latter phases of DWD evolution (i.e., right before the tidal disruption), the frequency and strain is enhanced within a very short time (e.g., several orbital periods), they can be treated as burst sources. These systems produce large amplitude signals that could be much higher than the detector noise and specific waveform models are not necessary in signal detection. In practice, the signal amplitude described for burst sources is the root-sum-square amplitude
\begin{align}
    h_\mathrm{rss}^2 = \int \mathrm{d}t\left(\left|h_+(t)\right|^2+\left|h_\times(t)\right|^2\right).
\end{align}
It can be approximated to $h_\mathrm{rss}\simeq|\tilde{h}(f)|\sqrt{\Delta f}$ by assuming that the GW mode is linearly polarized and $|\tilde{h}(f)|$ is almost constant within a frequency band $\Delta f$, where $\tilde{h}(f)$ is the Fourier transform of $h(t)$. However, the quantity $h_\mathrm{rss}$ is on the same order as the characteristic strain $h_c$ as defined before~(\citealt{Moore2014}). In this work, the main phase we focused on is the last stages of inspiraling before tidal disruption, and we therefore use $h_c$ to represent GW emission.

\subsection{Neutrinos}
\label{subsec:neu}

\subsubsection{neutrinos production}
We now consider neutrino emission from the merger event. When the density and temperature are high enough, thermal neutrino processes begin and produce $\mathcal{O}(1)\ \mathrm{MeV}$ neutrinos, at the same time neutrino cooling becomes efficient. In general, five thermal neutrino processes are taken into account: electron–positron annihilation, plasmon decay, photoemission, neutrino Bremsstrahlung and recombination, which produce neutrinos with all kinds of flavors~(\citealt{Itoh1996}). Once the temperature exceeds $10^9\  \mathrm{K}$, electron–positron annihilation
\begin{align}
    e^{+} + e^{-} & \rightarrow \nu + \bar{\nu},
\end{align}
starts to dominate the thermal neutrino production~(\citealt{Itoh1996}).

When the merging process is violent, the temperature can easily reach $\geq 5\times10^{9}\ \mathrm{K}$ such that weak processes including electron and positron capture becomes more efficient than thermal neutrino production channels. These weak processes mainly produce electron neutrinos and anti-electron neutrinos via,
\begin{align}
    p+e^{-} & \rightarrow n + \nu_e, \label{eqn:pe} \\
    n+e^{+} & \rightarrow p + \bar{\nu}_e, \\
    (Z,A) + e^{-} & \rightarrow (Z-1,A) + \nu_e \label{eqn:ZAe}, \\
    (Z-1,A) + e^{+} & \rightarrow (Z,A) + \bar{\nu}_e.
\end{align}
During the DWDs merging process, the material is proton-rich so that the electron capture happens mainly between free protons/nuclei and electrons, resulting in a dominant $\nu_e$ emission.

We assume that neutrinos follow the Fermi-Dirac distribution with zero chemical potential, and the neutrino number emitted per energy per time per area could be written as
\begin{align}\label{eqn:flux}
    \Phi_\nu(E) = \frac{L_{\nu}}{4\pi d^2}\frac{120}{7\pi^2}\frac{E^2}{T^4}\frac{1}{1+e^{E/k_B T}},
\end{align}
where $L_{\nu}$ is the neutrino luminosity, $T$ the neutrino temperature, $E$ the neutrino energy and $d$ the distance to the earth. Here we take the average neutrino energy to be $\left<E\right> \sim 3.15k_BT$, a result from the Fermi-Dirac distribution.

Neutrino luminosity and average energy strongly depend on the merging processes. Dynamical interactions of DWD failing to ignite can result in  optimistic values of neutrino luminosity with the total energy $\sim10^{48}\ \mathrm{erg}$~(\citealt{Aznar-Sigu'an2014}). As mentioned before, SNe Ia can produce a large amount of neutrinos~(\citealp{Kalogera2001, Wright2017}), so it should be reasonable to assume that SNe Ia from the double-degenerate channel can also generate a neutrino burst with the luminosity $\sim10^{49}\ \mathrm{erg\ s^{-1}}$ or even $\sim10^{50}\ \mathrm{erg\ s^{-1}}$. Here we adopt three characteristic temperature values in our study:
\begin{itemize}
    \item $\geq10^9\ \mathrm{K}$: this is above the carbon burning temperature, electron-positron pair annihilation starts to dominate the thermal production process;
    \item around $4\times10^9\ \mathrm{K}$: nuclear statistical equilibrium (NSE) starts, and the electron capture process become efficient;
    \item around $10^{10}\ \mathrm{K}$: it is the typical core neutrino temperature in a SN Ia explosion, resulting in the neutrino average energy $\left<E\right> \sim 3\ \mathrm{MeV}$.
\end{itemize}
During the neutrino burst, the weak or thermal production processes happen. For weak processes, the timescale of NSE can be approximated as $\tau_\mathrm{NSE}\sim 1\ \mathrm{s}$ for SNe Ia explosion~(\citealp{Kunugise2007}), which is the same order as the explosion timescale. Assuming that DWDs merging could also reach the NSE condition (i.e., cases 1, 2, 3(a), and 3(b)), it would be reasonable to assume that it also lasts for about 1 second. While in the thermal production, which is related to density and temperature during the merge processes, we assume that it lasts $\sim 1\ \mathrm{s}$, to be consistent with the case of explosion. The neutrino luminosity $L_\nu$, emission duration $\Delta t$ and temperature $T$ are assumed in this work as shown in Table \ref{tab:nu_cases}. In Table \ref{tab:nu_cases}, we argue that Cases 1 and 2 likely represent situations which will lead to SNe Ia, and neutrinos carry away $\sim(1-10)\%$ of the total released energy of $10^{51}\ \mathrm{erg}$. For Cases 3(a), 3(b) and 3(c), the luminosity is lower compared with the SNe Ia case, and we consider cases with temperature values (i.e., $4\times10^9\ \mathrm{K}$ or even $10^{10}\ \mathrm{K}$) which are somewhat optimistic.

\begin{table}
\bc
\begin{minipage}[]{100mm}
\caption[]{The neutrino luminosity, duration time and temperature we adopt in this work\label{tab:nu_cases}}
\end{minipage}
\setlength{\tabcolsep}{1pt}
\small
    \begin{tabular}{lcccc}
        \hline\noalign{\smallskip}
        {Cases} & {$L_{\nu_e}\ [\mathrm{erg\ s^{-1}}]$} & {$\Delta t\ [\mathrm{s}]$} & {$T \ [\mathrm{K}]$} & {Reference}   \\
        \hline\noalign{\smallskip}
        Case 1  & $10^{50}$ & $1$ & $\left<E\right>=3\ \mathrm{MeV}$ & (\citealt{Kunugise2007})\\
        Case 2  & $10^{49}$ & $2$ & $4\times10^{9}$ & (\citealt{Wright2017}) \\
        \hline\noalign{\smallskip}
        Case 3(a)  & $10^{48}$ & $1$ & $10^{10}$  & {   }  \\
        Case 3(b)  & $10^{48}$ & $1$ & $4\times10^{9}$  & (\citealt{Aznar-Sigu'an2014}) \\
        Case 3(c)  & $10^{48}$ & $1$ & $10^{9}$  & {   } \\
        \hline\noalign{\smallskip} 
    \end{tabular}
\ec
\end{table}

\begin{figure}[ht!]
    \centering
    \includegraphics[width=\hsize]{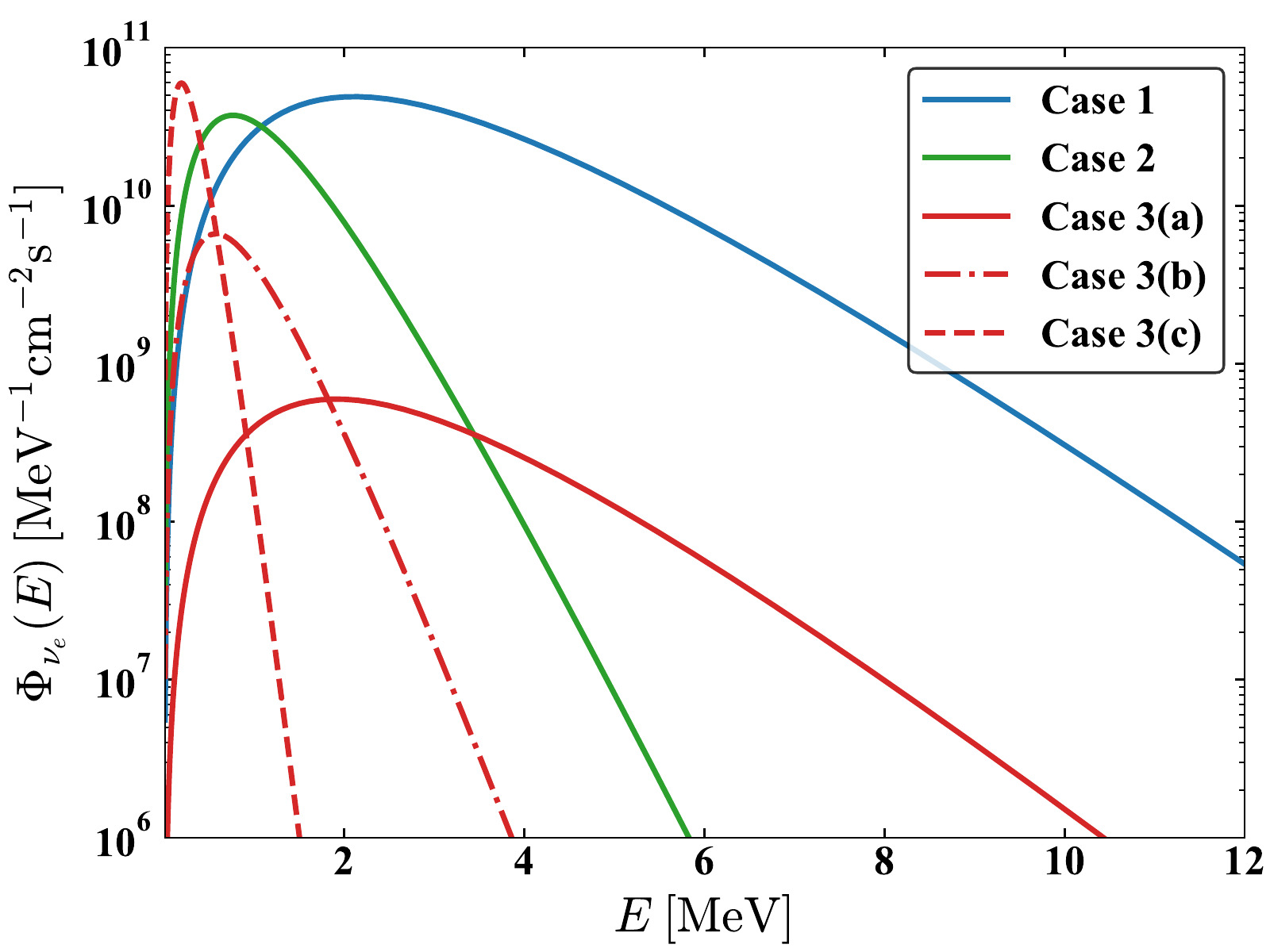}
    \caption{Neutrino fluxes of five cases presented in Table \ref{tab:nu_cases}, and the distance from the source is $d = 1\ \mathrm{kpc}$.}
    \label{fig:flux}
\end{figure}

Based on the assumed parameter values as in Table~\ref{tab:nu_cases}, the respective neutrino flux values are calculated with Eq. (\ref{eqn:flux}). From Figure \ref{fig:flux}, for higher temperature (averaged energy), the distribution of neutrino (flux) has broader energy range. 
Here, the material is proton-rich (electron fraction $Y_e>0.5$) during DWD merge, it implies that electron capture dominates neutrino production process, and $\nu_e$ will be dominant during the emission. \footnote{While in DNS merge $\bar{\nu}_e$ dominates due to positron capture in neutron-rich matters.} Nevertheless, it is a neutronization process which losses lepton number by neutrino release (nucleosynthesis that produce heavier elements also occurs). On the other hand, if the thermal burning process dominates, neutrinos of all flavors will be produced with similar proportion. However, if weak processes become significant then the production rate of thermal processes become sub-dominant at the same physical conditions, thus the thermal contribution to the neutrino flux can be safely ignored. We note also that no neutrino trapping is expected as occurred in core-collapse supernovae, because the matter density of DWD merger is not high enough\footnote{Core-collapse supernovae are triggered by massive star collapse. Before the explosion, the center core of star is so dense ($\geq10^{11}\ \mathrm{g\ cm^{-3}}$) that neutrinos cannot escape and are trapped in the inner regions for some time. Only after the explosion starts, the core density decreases and the trapping conditions do not exist any more, leading to a neutrino burst.}. 

\subsubsection{neutrino oscillation effects}

From the production region to surface, neutrino flavors will be regulated by the Mikheyev-Smirnov-Wolfenstein (MSW) effect~(\citealt{Wolfenstein1978,Mikheyev1985,Mattias2013}). These regulations are different depending on the mass order hierarchy: the normal mass hierarchy (NH) or the inverted mass hierarchy (IH)~(\citealt{Dighe2000,Fogli2005}). If we denote the initial flux as $\Phi_{\nu_\alpha}^0$ ( $\Phi_{\bar\nu_\alpha}^0$) where $\alpha = e, \mu, \tau$, then with MSW resonance effect, the flux at the surface is $\Phi_{\nu_\alpha}$ ($\Phi_{\bar\nu_\alpha}$). When neutrinos propagate from the source surface to the Earth, neutrino oscillation in vacuum has to be taken into account. The neutrino fluxes observed $\phi_{\nu_\alpha}$ ($\phi_{\bar\nu_\alpha}$) are different from $\Phi_{\nu_\alpha}$ ( $\Phi_{\bar\nu_\alpha}$). In general, the transition probabilities between $\phi_{\nu_\alpha}$ ( $\phi_{\bar\nu_\alpha}$) and $\Phi_{\nu_\alpha}$ ( $\Phi_{\bar\nu_\alpha}$) is:
\begin{align}
    \phi_{\nu_\beta}(E) = \sum_\alpha \Phi_{\nu_\alpha}(E)P_{\nu_\alpha\rightarrow\nu_\beta}\left({L}/{E}\right),\\ 
    \phi_{\bar\nu_\beta}(E) = \sum_\alpha \Phi_{\bar\nu_\alpha}(E)P_{\bar\nu_\alpha\rightarrow\bar\nu_\beta}\left({L}/{E}\right),
\end{align}
i.e., it depends on the energy of the neutrinos  and their travel distance (evolution time). However, the astrophysical sources are so far way from the Earth, so the probabilities should be averaged along the way and can be described as
\begin{align}
    P_{\alpha\beta} &= \left<P_{\nu_\alpha\rightarrow\nu_\beta}\right> = \sum_i^3 \left|U_{\alpha i}\right|^2\left|U_{\beta i}\right|^2, \\
    \bar{P}_{\alpha\beta} &= \left<P_{\bar\nu_\alpha\rightarrow\bar\nu_\beta}\right> = P_{\alpha\beta},
\end{align}
where $U_{\alpha i}$ is the element of Pontecorvo-Maki-Nakagawa-Sakata (PMNS) mixing matrix~(\citealt{Pontecorvo1957,Maki1962})
\begin{eqnarray}
 U  =  \left(
\begin{array}{ccc}
c_{12}c_{13} & s_{12}c_{13} & s_{13} e^{-i\delta} \\
-s_{12}c_{23}-c_{12}s_{23}s_{13}e^{i\delta} & c_{12}c_{23}-s_{12}s_{23}s_{13}e^{i\delta} & s_{23}c_{13} \\ 
s_{12}s_{23}-c_{12}c_{23}s_{13}e^{i\delta} & -c_{12}s_{23}-s_{12}c_{23}s_{13}e^{i\delta} & c_{23}c_{13}
\end{array}\right),
\end{eqnarray}
where $s_{ij} = \sin\theta_{ij}$, $c_{ij} = \cos\theta_{ij}$, in which $\theta_{ij}$ is the lepton flavor mixing angle and $\delta$ is the CP violation phase. The above parameters are taken from~\citet{Zyla2020}.

For the situation of electron neutrino dominant, $\Phi_{\nu_e}^0:\Phi_{\nu_x}^0:\Phi_{\bar\nu_e}^0:\Phi_{\bar\nu_x}^0 = 1:0:0:0$ , it leads to $\phi_{\nu_e}:\phi_{\nu_x}:\phi_{\bar\nu_e}:\phi_{\bar\nu_x} = 0.23:0.77:0:0$ for NH case and $0.32:0.68:0:0$ for IH case. Here, the subscript $e$ and $x$ denote the electron flavor and other ($\mu$ and $\tau$) flavors, and the bar denote the antineutrinos with different types. If three flavors of neutrinos are equally emitted, i.e. $\Phi_{\nu_e}^0:\Phi_{\nu_x}^0:\Phi_{\bar\nu_e}^0:\Phi_{\bar\nu_x}^0 = 1/6:1/3:1/6:1/3$, it has trivial effect $\phi_{\nu_e}:\phi_{\nu_x}:\phi_{\bar\nu_e}:\phi_{\bar\nu_x} = 1/6:1/3:1/6:1/3$. In this case, the neutrino mass order hierarchy cannot be distinguished.

\section{Detectability}
\label{sec:det}

\subsection{Gravitational waves}
The power of GW emission reaches the maximum when tidal disruption occurs and is largely suppressed when the secondary mass starts to be disrupted. We assume that the merging process occurs when the orbital separation $a_m$ is equal to the tidal disruption radius:
\begin{align}
    a_m = R_\mathrm{td} = qR_2
\end{align}
where $q$, $R_2$ is the mass-ratio ($m_2/m_1$) and the radius of the secondary WD, respectively. We assume the WD radius to be:
\begin{align}
    R \approx \left(\frac{3m}{4\pi\rho}\right)^{\frac{1}{3}},
\end{align}
where $\rho$ is the mean density of WD. The final cut-off GW frequency is then determined by Eq. (\ref{eqn:f_orb}). 

In this work, we consider four hypothetical DWD systems:  $(1.0+0.8)M_\odot$, $(1.0+1.0)M_\odot$, $(1.1+0.9)M_\odot$ and $(1.2+1.0)M_\odot$ which are usually assumed in simulations and might mostly lead to SNe Ia, which also have a high probability to produce neutrinos with significant luminosity. For various masses, the mean densities are taken to be $10^6\ \mathrm{g\ cm^{-3}}$ (for $0.8\ M_\odot$), $2\times10^6\ \mathrm{g\ cm^{-3}}$ (for $0.9\ M_\odot$) and $3\times10^6\ \mathrm{g\ cm^{-3}}$ (for $1.0\ M_\odot$), which we refer to Figure 1 of~\citet{Zou2020}.

Here, we use the Python package LEGWORK\footnote{\url{https://legwork.readthedocs.io/en/latest/}}~(\citealt{Wagg+2021}) to calculate the strain amplitude $h_{0}$ and characteristic strain amplitude $h_{c}$ of binaries systems and the evolution of their orbital frequency $f_\mathrm{orb}$. LEGWORK is also used to calculate SNRs of GW emitted from inspiraling binary systems as detected by certain GW detectors. However, in this work, we mainly focus on the latter stage of DWD merging process. We also take the initial GW frequency to be $5\ \mathrm{mHz}$ for four systems, and follow the evolution of GW frequency based on the two-point approximation as described in Sec. \ref{subsec:GW}.

\begin{figure}[ht!]
    \centering
    \includegraphics[width=\hsize]{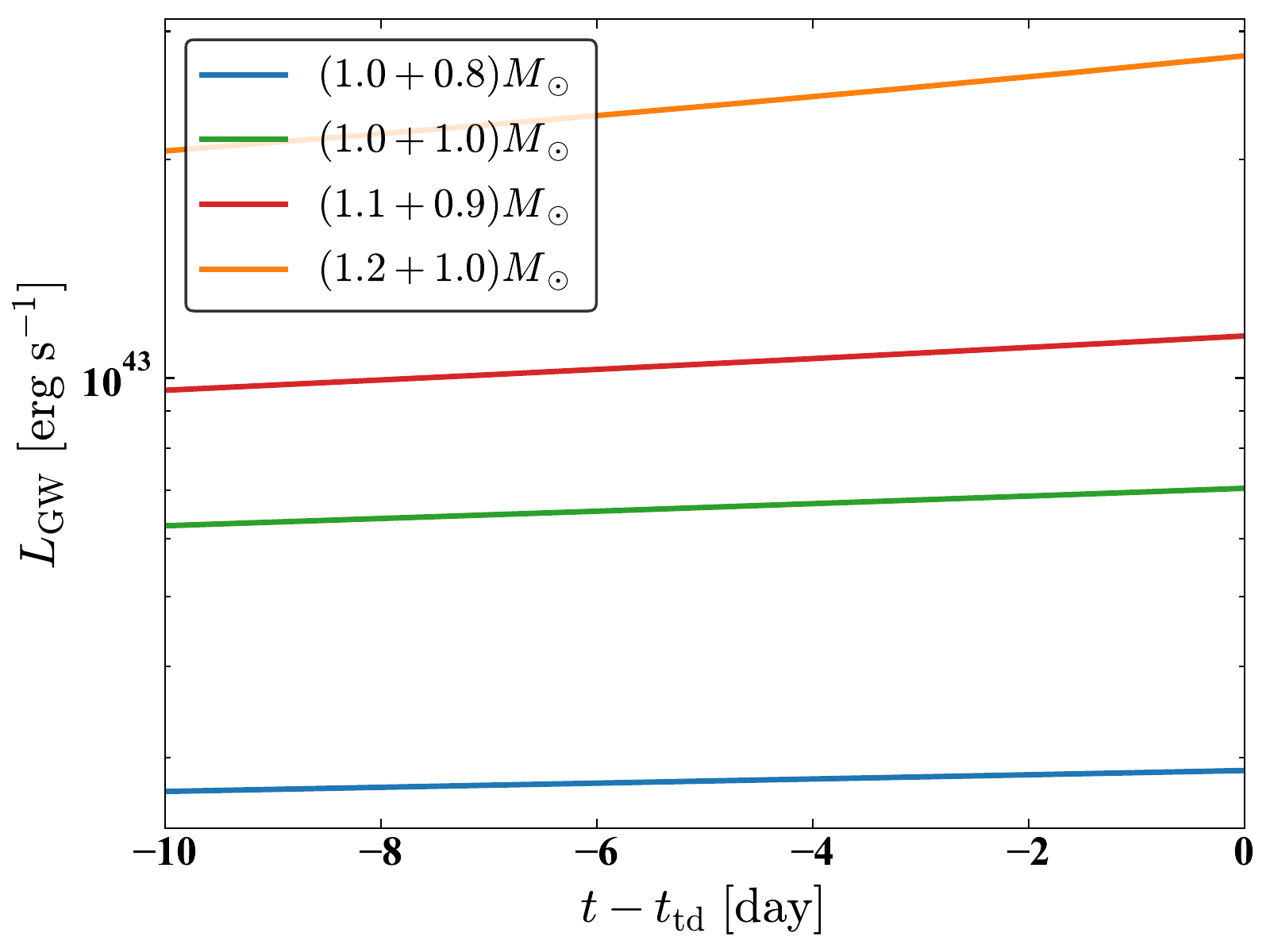}
    \caption{The GW luminosity in final evolution phase (i.e., beginning $10\ \mathrm{days}$ before the tidal disruption). The line colors denote systems with different component masses: $(1.0+0.8)M_\odot$, $(1.0+1.0)M_\odot$, $(1.1+0.9)M_\odot$ and $(1.2+1.0)M_\odot$.}
    \label{fig:L_GW}
\end{figure}

Figure \ref{fig:L_GW} show the luminosities of GW emission $L_\mathrm{GW}$ for these systems from $10$ days before tidal disruption,  $t_\mathrm{td}$, to the time of disruption. It can be seen that for more massive systems, the luminosities are higher as expected. With the same total mass, the equal mass-ratio (e.g., $(1.0+1.0)M_\odot$) system has a smaller luminosity than the unequal mass-ratio (e.g., $(1.1+0.9)M_\odot$) one. Integrating luminosities from $t_\mathrm{td}-10\ \mathrm{day}$ to $t_\mathrm{td}$, it follows that the energy release by GW emission is on the order of $10^{48}-10^{49}\ \mathrm{erg}$.

\begin{figure}[ht!]
    \centering
    \includegraphics[width=\hsize]{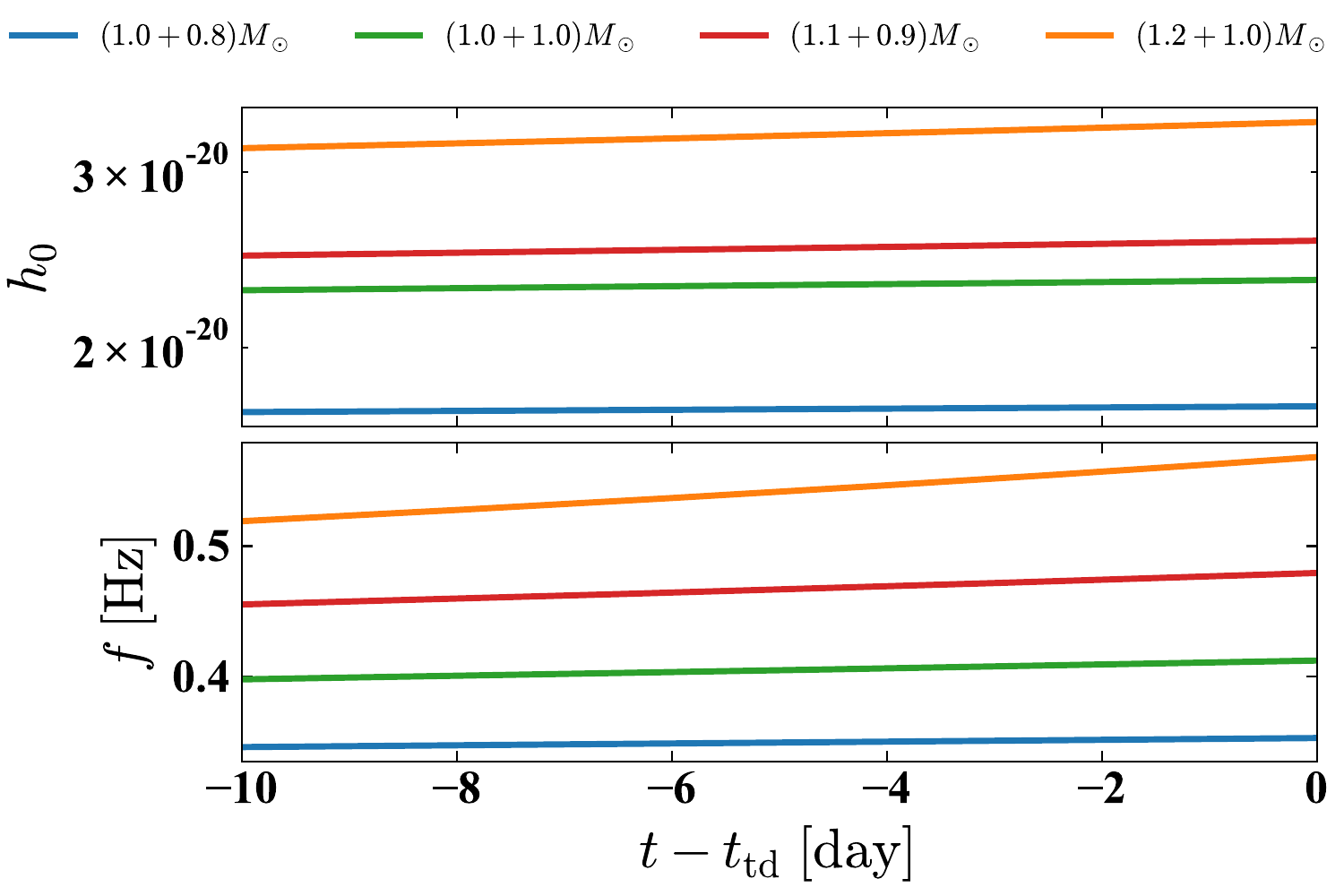}
    \caption{The final evolution of strain amplitude $h_0$ and GW frequency $f$ beginning $10$ days before the tidal disruption. The line colors represent the same systems as in Fig.~\ref{fig:L_GW}.}
    \label{fig:evol}
\end{figure}

We assume the distance from source to the Earth to be $1\ \mathrm{kpc}$, and calculate $h_0$ and $h_c$ which could represent the instantaneous moment and \emph{accumulated} strength of GW emission. We show the evolution of $h_0$ and GW frequency of these systems since $10\ \mathrm{days}$ before tidal disruption in Figure \ref{fig:evol}. It can be seen that $h_0$ and GW frequency are not evolving with the same rate of change, i.e., the latter is changing more rapidly. It implies that the strain amplitude varies across the frequency band in the final stage of merging. Besides, the $h_0$ and $f$ of DWDs are \emph{slightly} increasing whereas in the cases of DNSs and DBHs they change suddenly. This contrast can be understood because the components of binaries are more compact and have smaller radii in the latter cases. Thus, the cut-off frequencies and the evolution rates of late stage are larger. On the other hand, the merge of DWD seems gentle comparing with DNSs and DBHs. Nevertheless, the GW signals will be detected by planned space-based interferometers assuming a distance of $1\ \mathrm{kpc}$.

\begin{figure}[ht!]
    \centering
    \includegraphics[width=\hsize]{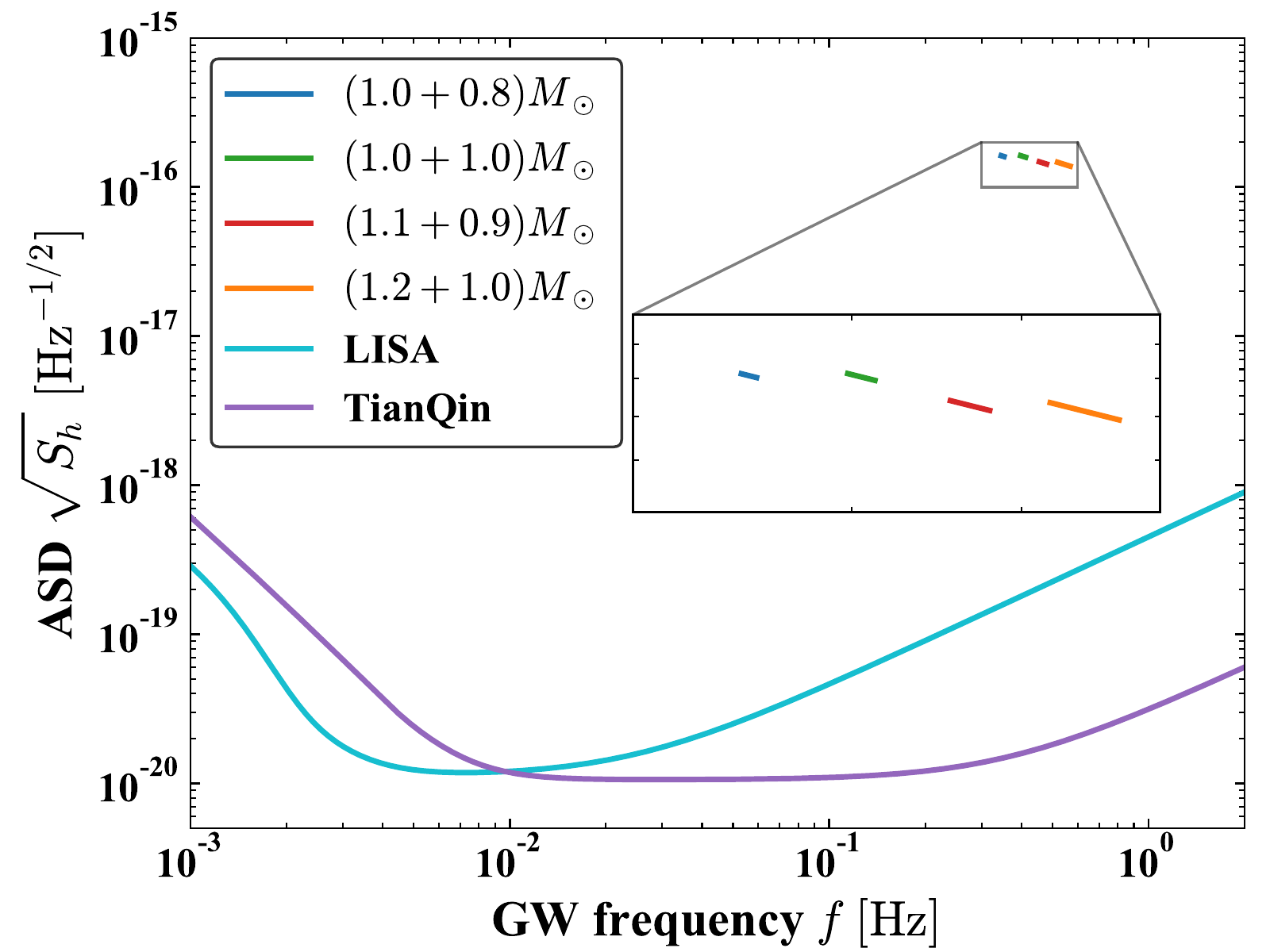}
    \caption{The GW amplitude spectral density (ASD) of the four systems as presented in Fig.~\ref{fig:L_GW}. The sensitivity curves of LISA and TianQin are also showed in this figure. GW frequency in these cases are in $0.3-0.6\ \mathrm{Hz}$ band which can be well detected by space-based GW detectors. Here we assume the source is $1\ \mathrm{kpc}$ away from the Earth.}
    \label{fig:ASD}
\end{figure}

Concerning the detectability of the GW signal, it is a common practise to compare the ASD $\sqrt{S_h}$ with the sensitivity curves of detectors where $S_h$ represents the energy density. Because we are only concerned about the final stage of inspiraling, we can safely ignore the galactic confusion noise which is related to detector's observing time. From Figure \ref{fig:ASD}, if the sources is at the distance of $1\ \mathrm{kpc}$, it is clear that the four systems we assume have a significant signal strength that is easily detected by LISA~(\citealt{Robson2019}) and TianQin~(\citealt{Huang2020}) in $0.3-0.6\ \mathrm{Hz}$ frequency band within final $10$ days before tidal disruption. We do not calculate the related SNRs because they are so large and the noise is unimportant due to the short time duration in our consideration.

\subsection{Neutrinos}

Neutrinos from DWD merger events are in the $\mathrm{MeV}$ energy scale and are observable via inverse beta decay (IBD) and electron-neutrino elastic scattering (ES) channels. The IBD channel is typically used to detect $\bar{\nu}_e$, while the ES channel is used to detect $\nu_e$ and other types of neutrinos. The threshold energy of the IBD channel is $\sim1.8\ \mathrm{MeV}$, while that of the ES channel heavily depends on the material used in the detectors. For water Cherenkov detectors such as Super-Kamiokande (Super-K)~(\citealt{Super-K}) and Hyper-Kamiokande (Hyper-K)~(\citealt{Hyper-K}), the threshold energy of recoil electrons is $\sim0.77\ \mathrm{MeV}$, corresponding to a minimum neutrino energy of $\sim0.97\ \mathrm{MeV}$. The fiducial volume of the inner detector of Super-K and Hyper-K is $22.5\ \mathrm{kt}$ and $220\ \mathrm{kt}$, respectively. The corresponding target particle number is $N_\mathrm{tar}\approx(M_\mathrm{tar}/m_p)\times(2/18)=6.7\times10^{30}(M_\mathrm{tar}/ 1\ \mathrm{kt})$, where $M_\mathrm{tar}$ is the effective mass of water and $m_p$ is the proton mass. Therefore, the target electron number of Super-K and Hyper-K are, respectively, $1.51\times10^{32}$ and $1.47\times10^{33}$. For JUNO -- a liquid organic scintillation, the target number is $1.29\times10^{33}$~(\citealt{JUNO:An2016}) with a rather low detection energy threshold of $0.2\ \mathrm{MeV}$~(\citealt{JUNO:Fang2020}) that corresponds to the neutrino threshold energy of $0.35\ \mathrm{MeV}$. The cross-section of IBD and ES interaction channels are taken from the code package provided by~\citet{SNOwGLoBES}, and the formulae of IBD and ES are taken from~\citet{bib_IBD,bib_ES}. In this work, $\nu_e$ dominate the emission, so we mainly focus on the ES channel.

The neutrino event number can be calculated with
\begin{align}
    N_\nu = N_\mathrm{tar}\Delta t\int^{E_\mathrm{max}}_{E_\mathrm{th}}\phi(E)\sigma(E)\eta(E)\mathrm{d}E,
\end{align}
where $N_\mathrm{tar}$ is the target particle number of protons, $\Delta t$ denotes the event duration ( observed time window), $\sigma(E)$ is the interaction-channel cross section, $\phi(E)$ is neutrino fluxes at earth, $\eta(E)$ is the efficiency of neutrino detector and is usually taken as $\eta\simeq0.9$.

We could calculate the neutrino numbers and show in Table \ref{tab:N_nu}.
\begin{table}
\bc
\begin{minipage}[]{100mm}
\caption[]{The electron neutrino $\nu_e$ (other neutrino $\nu_x$) events number calculated in detectors sensitive to $\mathcal{O}(1\ \mathrm{MeV})$. NH and IH denotes the normal and inverted mass order hierarchy for neutrino respectively. In this table, the values are presented in scientific notation and values in bracket are for ${\nu}_x$\label{tab:N_nu}}
\end{minipage}
\setlength{\tabcolsep}{1pt}
\small
    \begin{tabular}{ccccccc}
        \hline\noalign{\smallskip}
        {Detector} & {hierarchy} & {Case 1} & {Case 2} & {Case 3(a)} & {Case 3(b)} & {Case 3(c)} \\
        \hline\noalign{\smallskip}
        \multirow{4}{*}{JUNO}     & \multirow{2}{*}{NH}  & 1.2E-00 & 2.0E-01 & 1.2E-02 & 9.8E-03 & 1.6E-03 \\
                              &   & (7.2E-01)  & (1.3E-01)  & (7.2E-03)  & (6.7E-03)  & (1.2E-03)\\
                              & \multirow{2}{*}{IH}  & 1.6E-00 & 2.7E-01 & 1.6E-02 & 1.4E-02 & 2.2E-03\\
                              &   & (6.4E-01)  & (1.2E-01)  & (6.4E-03) & (6.0E-03) & (1.1E-03) \\
        \hline
        \multirow{4}{*}{Super-K}  & \multirow{2}{*}{NH}  & 1.4E-01 & 1.8E-02 & 1.3E-03 & 8.9E-04 & 5.0E-06 \\
                              &   & (8.4E-02) & (1.2E-02) & (8.3E-04) & (6.0E-04) & (3.5E-06)\\
                              & \multirow{2}{*}{IH}  & 1.9E-01 & 2.5E-02 & 1.9E-03 & 1.2E-03 & 7.0E-06\\
                              &   & (7.4E-02) & (1.1E-02) & (7.3E-04) & (5.3E-04) & (3.1E-06)\\
        \hline
        \multirow{4}{*}{Hyper-K}  & \multirow{2}{*}{NH}  & 1.3E-00 & 1.7E-01 & 1.3E-02 & 8.7E-03 & 4.9E-05\\
                              &   & (8.2E-01) & (1.2E-01) & (8.1E-03) & (5.8E-03) & (3.4E-05)\\
                              & \multirow{2}{*}{IH}  & 1.8E-00 & 2.4E-01 & 1.8E-02 & 1.2E-02 & 6.8E-05 \\
                              &   & (7.2E-01) & (1.0E-01) & (7.2E-03) & (5.1E-03) & (3.0E-05)   \\
        \noalign{\smallskip}\hline
    \end{tabular}
\ec
\end{table}
From Table \ref{tab:N_nu}, one can see that in the most optimistic situation (Case 1), the neutrino number $N_{\nu_e}$ ($N_{\nu_x}$) can reach about $2$ (at the order of $\mathcal{O}(1)$). $N_\nu$ is different between NH and IH, i.e., the ratio of NH and IH ($N_\mathrm{NH}/N_\mathrm{IH}$) is $0.72$ for $\nu_e$ and $1.13$ for $\nu_x$. The neutrino event number of Case 3(c) is very small because the average energy of this case are around the threshold of water Cherenkov detectors. It is clear that the higher temperature of neutrinos, the more events we observe in detectors. Furthermore, if the evolution of luminosity is specified, one could expect the event numbers evolves with time, which reflect the physics of merge processes, particularly the explosion mechanism if it leads to SN Ia. However, these processes and mechanism are still poorly studied, and we simply take the average of luminosity with $\Delta t \sim1\ \mathrm{s}$ time window. Nevertheless, it means that the neutrino production processes are important ingredients in understanding the merging physics.

\section{Discussion and Conclusions}
\label{sec:dis&con}

In this work, we show that DWD merger events are able to produce detectable GW and neutrino signals if they are located at distance $1\ \mathrm{kpc}$. From GW point of view, the GW frequency DWD systems are within $0.3-0.6\ \mathrm{Hz}$ band in $10$ days before the tidal disruption in the four example DWD systems that we have studied. Unlike DNSs and DBHs, which are the main potential sources of ground-based GW detectors such as the Advanced LIGO and Virgo, DWD systems are more easily to be detected by space-based GW detectors such as LISA and TianQin at the last stage of inspiraling. Besides, the evolution of strains and GW frequency are moderate while for DNSs and DBHs they are more dramatic. This can be understood since WD are not so dense and the GW signals reach the maximum at tidal disruption radius and thereafter the two bodies merge. The amplitude spectral density of these systems are large comparing to the sensitive curves of LISA and TianQin for the final short time interval, implying that DWDs merging could be well detected by space-based GW detectors.

From neutrino point of view, given that the average energy from these events is $\mathcal{O}(1)\ \mathrm{MeV}$, the upper limit of neutrino event numbers is on the order of $\mathcal{O}(1)$ for current and upcoming detectors such as Super-K, Hyper-K and JUNO. We found that the merger remnant cases (i.e., case 3(a), 3(b) and 3(c)) are harder to detect because of the lower neutrino luminosity. In $\mathcal{O}(1)\ \mathrm{MeV}$ energy, the background mainly come from solar neutrinos but it can be easily removed given the short duration time. We also show that the detected neutrino flux could help to distinguish the neutrino mass order hierarchy when taking the neutrino oscillation into account. Therefore, we need detectors with larger volume (i.e., more target number) and lower energy threshold, which will also extend the detection horizon distance. More theoretical studies about merging physics and explosion mechanisms of DWDs are also necessary to provide the neutrino luminosity, spectra and oscillation mode.

In the era of multi-messenger astronomy, the DWD merging events are important sources and can be studied using various  multi-messenger emissions. Apart from GW and neutrino signals, the subsequent electromagnetic signals such as optical-infrared, X-rays and bolometric light curves are also interesting and important observable signatures~(\citealp{Rueda2019,Moll2014,Aznar-Sigu'an2014}). Even if the electromagnetic signal could be similar between different explosion mechanisms (e.g., DD and SD SNe Ia) and/or merging processes, the neutrino detection can help to distinguish theoretical scenarios once the neutrino production models of DWDs merging are specified. The detection of these sources can join the next-generation supernova early warning system (SNEWS 2.0)~(\citealp{SNEWS2.0}), and will guide the observation of DWD merging events in the future.

\normalem
\begin{acknowledgements}
This work is supported by the National Natural Science Foundation of China (NSFC) grants 11633007, 12005313 and U1731136, Guangdong Major Project of Basic and Applied Basic Research (Grant No. 2019B030302001), Key Laboratory of TianQin Project (Sun Yat-sen University) of the Ministry of Education, and the China Manned Space Project (No. CMS-CSST-2021-B09).
\end{acknowledgements}

\bibliographystyle{raa}
\bibliography{refs}

\end{document}